\documentclass[12pt]{article}
\usepackage{amssymb,amsmath,amsthm,amsfonts,amscd}
\usepackage{graphicx}
\date{\small}
\newcommand\be{\begin{equation}}
\newcommand\ee{\end{equation}}
\newcommand\bea{\begin{eqnarray}}
\newcommand\eea{\end{eqnarray}}

\newcommand{\fatalpha}{{\bf \alpha \kern -0.44em \alpha}}
\newcommand{\fatsigma}{{\bf \sigma \kern -0.54em \sigma}}
\newcommand{\tpchi}{{\bf D \kern -0.35em D}}
\newcommand{\llambda}{{\bf \lambda \kern -0.45em \lambda}}

\renewcommand{\theequation}{\arabic{equation}}
\renewcommand{\theequation}{\thesection.\arabic{equation}}
\bibliography{plain}
\title{\bf \large{ A general algorithm for manipulating non-linear and linear entanglement
witnesses by using  exact convex optimization}} \vspace{20mm}
\author{ M. A. Jafarizadeh$^{a,b,c}$
\thanks{E-mail:jafarizadeh@tabrizu.ac.ir},
K. Aghayar$^{a}$ \thanks{E-mail:Aghayari@tabrizu.ac.ir}, A.
Heshmati$^{a}$ \thanks{E-mail:Heshmati@tabrizu.ac.ir}
 \\
$^a${\small Department of Theoretical Physics and Astrophysics,
University of Tabriz, Tabriz 51664, Iran.}  \\ $^b${\small
Institute for Studies in Theoretical Physics and Mathematics,
Tehran 19395-1795, Iran.}\\$^c${\small Research Institute for
Fundamental Sciences, Tabriz 51664, Iran.}} \pagebreak
\pagebreak

\vspace{20mm}
\begin{document}
\maketitle \vspace{15mm}

\begin{abstract}
A generic algorithm is developed to reduce the problem of obtaining linear and nonlinear entanglement witnesses of a given quantum system, to convex optimization problem. This approach is completely general and can be applied   for  the entanglement detection of any N-partite quantum system. For this purpose, a map from convex space of separable density matrices to a convex region called feasible region is defined, where by using exact  convex optimization method, the linear entanglement witnesses can be obtained  from polygonal shape feasible regions, while for curved shape feasible regions, envelope of the family of linear entanglement witnesses can be considered as nonlinear entanglement witnesses. This method proposes a new methodological framework within which most of previous EWs can be studied. To conclude and in order to demonstrate the capability of the proposed approach,  besides providing  some nonlinear witnesses for entanglement detection of density matrices in unextendible product bases, W-states, and GHZ with W-states, some further examples of three qubits systems and their classification and entanglement detection are included. Also it is explained how one can manipulate most of the non-decomposable linear and nonlinear three qubits entanglement witnesses  appearing in some of the papers published by us and other authors, by the method proposed in this paper.

\end{abstract}
\hspace{1cm}{\bf Keywords: }
non-linear and linear entanglement witnesses, convex optimization\\
\hspace{1cm}{\bf PACS number(s): 03.67.Mn, 03.65.Ud}

\section{Introduction}
\par
Entanglement is one of the interesting features of quantum systems. It is used as a physical
resource in realization of many quantum information and quantum computation processes such
as quantum parallelism \cite{Deutsch}, quantum cryptography \cite{Ekert}, quantum teleportation \cite{Bennett,Bouwmeester}, quantum dense coding \cite{Wiesner,Mattle}, reduction of communication complexity \cite{Cleve} and beating classical communication complexity bounds with entanglement \cite{HorodeckiE}. In these applications usually a source produces entangled particles and after these particles reach to the related parties, there is an important question for the parties - are these particles already entangled?
\par
One approach to distinguish entangled states from separable ones is entanglement witness (EW) \cite{HorodeckiE,terhal}. A quantum state is entangled iff there exists a Hermitian operator W with $Tr(W \rho) < 0$ and $Tr(W \rho_{sep}) \geqslant 0$ for any separable state $\rho_{sep}$ \cite{Bruss}. We say that the witness W detects the entanglement of density matrix. Recently there has been an increased interest in the nonlinear EWs because of their improved detection with respect to linear EWs. A nonlinear EW is any bound on nonlinear function of observables which is satisfied by separable states but violated by some entangled states \cite{HorodeckiE,Moroder,JMHA,Jaf8}.
\par
Optimization problems occur in both classical and quantum physics \cite{Hartmann}. One of the important subclass of optimization is convex optimization where the related functions of problem are convex. The importance of convex optimization is that in these optimizations, any locally optimal solution of problem is guaranteed globally optimal \cite{Chong}. On the other hand, the set of all possible states of a quantum system that can occur in nature must be a convex set \cite{Bengtsson} and as the state  of a quantum system is fully characterized by density matrix of that system so the density matrix must be a convex function. Therefore, convex optimization is a natural optimization in quantum information.
\par
Although all the work on this paper deals with obtaining EWs via convex optimization, other optimization approaches such as linear and semi-definite programming methods, can be found in the literature. For example, the reader can find obtaining some EWs by linear programming in \cite{JafRezSey,JafNajHab,JafAkbHab}, semi-definite programming for distinguish entangled from separable quantum states and using robust semi-definite programs and EWs to study the distillability of the Werner states in \cite{Doherty1,Vianna}, and convex optimization applications in entanglement in \cite{Auden1}.
\par
In this paper, we provide a general algorithm for finding the EWs by exact convex optimization method. For this purpose, for a given system or density matrix we determine the feasible region (FR). The FR for a system is defined by the mapping from separable states space to a region called feasible region i.e.  $Tr( W\rho_{s} )$ where $\rho_{s}$ is separable density matrices of that system. As the $\rho_{s}$ has convex structure, FR must be convex too ( the defined mapping do not change the convexity property ). Any tangent to the surface of this FR corresponds to an EW because it separates at least an entangled state from separable states. If this FR was a polygon, applying first convex optimization to this convex function, would give linear EWs which are one of vertices of polygon but if the FR was not a polygon, then applying convex optimization would give a family of linear EWs which are tangent to FR.
Nonlinear EW could be considered as the envelope of these family \cite{JAAHM}. The key point for convex optimization arise from the linear or nonlinear form of FRs therefore linear or nonlinear cost functions and constraints in convex optimization problem. Although we will not consider in this paper, if one can not determine the FR exactly, one can solve the problem by approximating the FR ( for example one can encircle the FR with a polygon \cite{JafRezSey}). Also, if one can not solve the convex optimization problem analytically, there are efficient numerical methods such as interior point method which may solve the problem numerically. After then we consider the entanglement detection problem of given density matrix with EWs in the previous part. The optimized EW(s) is come from the reapplication of convex optimization with new constraints. Although this method is general and could be applied for any quantum systems, here we present examples with some new EWs for three qubits systems.
\par
The structure of the article is as follows. Sec. II introduces FR for a given system and discusses how to determine FR for some selected operators in the Hilbert space of that system. In Sec. III convex optimization is applied for finding linear EWs using FR which is determined in second section. In Sec. IV convex optimization is applied again for finding nonlinear EWs using results of Sec. III. In Sec. V, we list some important linear and nonlinear EWs for three qubits systems which have been detected by convex optimization. The non-decomposability of these EWs are also discussed. Optimality of some EWs including linear EWs and a special case of spherical case is presented in Sec. VI. The detection of these EWs for some important three-qubits density matrices such as density matrices in unextendible product bases, W-state, and mixed GHZ with W states density matrices, have been presented in Sec. VII. Convex optimization review and some detailed proofs of paper would presented in appendices. Throughout these section, we have presented examples with details for three qubits system to present the practicality of this method.
\section{Feasible region}
\par
One of the main problems in quantum information processing is detecting the entanglement of the system. For a given state of a quantum system
i.e. density matrix, we want to find some (particularly optimal) EW's for detecting entanglement of the system.
\par
Consider a multipartite quantum system consisting of n subsystems which is characterized by density matrix. In real applications of quantum information density matrices are mixed. A mixed state of n systems is entangled if it cannot be written as a convex combination of product states \cite{HorodeckiE} $\rho\neq\sum_{_{i}} p_{i} \rho_{1}^{i}\otimes ... \otimes \rho_{n}^{i}$ with $p_{i} \geqslant 0$ and $\sum p_{i} = 1$, otherwise it is separable. The total Hilbert space H of n systems is a tensor product of the subsystem spaces $H =\otimes_{_{i=1}}^{^{n}} H_{_{i}} $ and any Hermitian operator such as EW could be written as a combination of operators $Q_{i}$ in this total Hilbert space.
\par
Now consider a set of Hermitian operators $Q_{_{i}}$. This set of operators are chosen in a way that the entanglement of the system could be detected. We will attempt to construct various linear and non-linear EWs using these operators. To this aim, for any separable state $\rho_{s}$ we introduce the maps
\begin{equation}\label{varp}
    P_{_{i}}=Tr(Q_{_{i}}\rho_{s})
\end{equation}
which map the convex set of separable states into a convex region named the feasible region (FR). Any hyper-plane tangent to the FR corresponds to a linear EW, since such hyper-planes separate the FR from entangled states. Hence, we need to determine the geometrical shape of FR. In general, determining the geometrical shape of FR is a difficult task. However, one may choose the Hermitian operators $Q_{_{i}}$ in such a way that the exact geometrical shape of FR can be obtained rather simply. By such a choice, when the FR is a polygon, its surface corresponds to
linear EWs which are linear combinations of the operators $Q_{_{i}}$; otherwise, linear EWs come from any hyper-plane tangent to the surface of FR. When the FR is not a polygon, there are a family of linear EWs which any of them are tangent to FR and the envelope of this family could be considered as nonlinear EW. \cite{Jaf8,JMHA}.
\par
In summary, there are two kinds of inequalities which determines the FR, nonlinear and linear ( see example 2 ). Therefore, the FR are constructed with some hyper-surfaces coming from nonlinear inequality constraints and some hyper-planes coming from linear inequality constraints. As any hyper surface tangent to FR is an EW, the envelope of linear EWs family each tangent to the nonlinear part of FR could be considered as nonlinear EW's. Finally, linear inequalities lead to linear EW's which are tangent to linear surfaces of FR.
\par
In analyzing the FR, there are three cases for region defined by nonlinear and linear inequality constraints. In first case, the region defined only by linear inequality constraints, i.e. $g_{i}(P_{1},..,P_{n})$, lie completely outside the region defined only by nonlinear inequality constraint, i.e. $f(P_{1},..,P_{n})$. In this case, the nonlinear constraints define the FR completely. In the second case, the region defined only by linear inequality constraints lie completely inside the region defined by only nonlinear inequality constraint. In this case, the linear constraints define the FR completely. And finally in the third case, the nonlinear and linear inequality constraints have some inter sections and due to nature of convex optimization, the optimal point in the FR is one of these intersection or lie in the intersections of linear constraints ( see following examples ).

\textbf{\emph{Example 1: FR with polygonal shape for three qubits systems}}
\par
As a special case we try to find FR with polygonal shape for a three qubits system. The operators in this Hilbert space could be written as tensor product of Pauli group operators for qubit i.e.
\begin{equation}\label{notat}
    \sigma_{i}\otimes\sigma_{j}\otimes\sigma_{k},\quad
    i,j,k=0,1,2,3.
\end{equation}
where $\sigma_{0}$, $\sigma_{1}$, $\sigma_{2}$ and $\sigma_{3}$ stand for the two dimensional identity operator $I_{2}$ and single qubit
Pauli operators $\sigma_{x}$, $\sigma_{y}$ and $\sigma_{z}$ respectively. For simplicity hereafter we will use the notation $I_{2}=I$, $\sigma_{x}=X$, $\sigma_{y}=Y$, and $\sigma_{z}=Z$ and will skip over the tensor product notation. The general task is to find linear and nonlinear relations between operators in the Hilbert space of three-qubits. As there are many ways for choosing set of operators and then finding relation between them, we may restrict ourselves to operators which appear in the expansion of given density matrix of system in terms of Pauli operators. As an example consider the following set of operators.
\begin{equation}\label{FRLinear}
    Q_{1}=XXX,\quad Q_{2}=XYY,\quad Q_{3}=YXZ,\quad Q_{4}=YZY,\quad Q_{5}=ZYZ,\quad Q_{6}=ZZX.
\end{equation}
The linear constraints are (see appendix B)
\begin{equation}\label{LiConst}
(-1)^{^{i_{1}}} P_{1}+(-1)^{^{i_{2}}} P_{2}+(-1)^{^{i_{3}}} P_{3}+ (-1)^{^{i_{4}}}P_{4}+(-1)^{^{i_{5}}} P_{5}+(-1)^{^{i_{1}+i_{2}+i_{3}+i_{4}+i_{5}+1}} P_{6} \leqslant 1
\end{equation}
and the FR is a polygon which its boundary planes are (\ref{LiConst}).\\
\textbf{\emph{Example 2: FR with quadratic and polygonal shape for three qubits systems}}
\par
This is an example that there are nonlinear constraints in addition to linear ones for FR. If we choose the following set of operators
$$
Q_{_{1}}=ZXX+ZYY,\quad
Q_{_{2}}=XXX+XYY,\quad
Q_{_{3}}=YXX+YYY,\quad
$$
\begin{equation}\label{Op}
    Q_{_{4}}=ZXY+ZYY,\quad
Q_{_{5}}=XXY+XYX,\quad
Q_{_{6}}=YXY+YYX,\quad
\end{equation}
\begin{equation}\label{OpE}
    Q_{_{7}}=IXZ,\quad
    Q_{_{8}}=IYZ,\quad
    Q_{_{9}}=IZI,\quad
\end{equation}
some trigonometric calculations (see Appendix B) lead to the following FR
\begin{equation}\label{e1}
    \sum_{i=1}^{9} P_{i}^{2} \leqslant 1
\end{equation}
which is a hyper ball in $P_{i}s$  space. In addition to this nonlinear hyper-surfaces, i.e. (\ref{e1}), there are some linear hyper-planes which restrict the FR. These are
$$\pm P_{1} \pm P_{5}\leqslant 1 , \quad \pm P_{1} \pm P_{6}\leqslant 1 , $$
$$\pm P_{2} \pm P_{4}\leqslant 1 , \quad \pm P_{2} \pm P_{6}\leqslant 1 , $$
\begin{equation}\label{const}
   \pm P_{3} \pm P_{4}\leqslant 1 , \quad \pm P_{3} \pm P_{5}\leqslant 1 .
\end{equation}
As in geometry, a spherical cap is a portion of a sphere cut off by a plane so one can say that the FR is a hyper-ball cap but now cut off by $24$ planes in (\ref{const}).

\section{Constructing linear EWs via convex optimization}
\par
After determining the FR which is a convex region, we can convert the problem of finding EWs to the convex optimization problem. Now we can construct EWs from operators $Q_{i}$ which have been used before for obtaining the FR. For this purpose consider a Hermitian operator W with some negative eigenvalues
\begin{equation}\label{w1}
    W=A_{0} I + \sum_{i} A_{i} Q_{i}
\end{equation}
where $A_{0}$ is nonzero positive real, $I$ is identity matrix with dimensionality equal with the Hilbert space of the system, $Q_{i}$ are positive operators with $-1 \leqslant Tr(Q_{i}\rho_{s}) \leqslant 1$, for every
separable states $\rho_{s}$, and $A_{i}$ are real parameters whose
ranges must be determined in a way that W become an EW.
\par
From definition of EW the condition, $Tr( W \rho_{s}) \geqslant 0$, must be satisfied. In order to satisfy this condition, we use convex optimization as follows. For ( \ref{w1} ), using convex optimization we can minimize the term
\begin{equation}\label{w2}
    Tr( W \rho_{s})=A_{0}+\sum_{i} A_{i} P_{i}
\end{equation}
where $A_{0}$ would be chosen in a way that $Tr( W \rho_{s}) \geqslant 0$. Although we apply this procedure for a pure state, but if the minimum of (\ref{w2}) is positive with all pure states, it will be positive for mixed states because mixed states could be written as convex combination of pure states.
\par
To summarize, the convex optimization problem takes the form
$$
\mathrm{minimize} \quad  A_{0}+\sum_{i}^{n} A_{i} P_{i}
$$
\begin{equation}\label{w3}
    \mathrm{subject\quad to} \quad  f(P_{1},..,P_{n})\leqslant 0,
\end{equation}
$$
g_{i}(P_{1},..,P_{n})\leqslant 0  \quad \mathrm{for} \quad i=1,...,m .
$$
where $f(P_{1},..,P_{n})$ is nonlinear inequality constraint and $g_{i}(P_{1},..,P_{n})$ are linear inequality constraints.


\emph{\textbf{Example 3: Linear EWs for three qubits system}} \label{exa3}
\par
As a special case, in this section we obtain linear EWs for three qubits system with some details. From FR obtained in example 2, we minimize the function $\sum_{i=1}^{9} A_{i} P_{i}$.
So our convex optimization problem takes form
\vspace{5mm}
$$
\mathrm{Minimize} \quad f(P_{1},..., P_{9} ) =\sum_{i=1}^{9}A_{i} P_{i}
$$
$$
\mathrm{subject \ to} \quad \sum_{i=1}^{9}P_{i}^{2}-1\leqslant0,
$$
$$
\pm P_{1} \pm P_{5}\leqslant 1 ,\quad \pm P_{1} \pm P_{6}\leqslant 1 ,
$$
$$
\pm P_{2} \pm P_{4}\leqslant 1 ,\quad \pm P_{2} \pm P_{6}\leqslant 1 ,
$$
$$
\pm P_{3} \pm P_{4}\leqslant 1 , \quad \pm P_{3} \pm P_{5}\leqslant 1 .
$$
The Lagrangian for this problem is
$$
\mathbf{L}(P, \lambda) =\sum_{i=1}^{9} A_{i} P_{i} + \lambda_{1}(\sum_{j=1}^{9}P_{j}^{2}-1)
$$
$$
+ \lambda_{2}( + P_{1} +  P_{5} - 1)+ \lambda_{3}( + P_{1} -  P_{5} - 1)+ \lambda_{4}( - P_{1} +  P_{5} - 1)+ \lambda_{5}( - P_{1} -  P_{5} - 1)
$$
$$
+ \lambda_{6}( + P_{1} +  P_{6} - 1)+ \lambda_{7}( + P_{1} -  P_{6} - 1)+ \lambda_{8}( - P_{1} +  P_{6} - 1)+ \lambda_{9}( - P_{1} -  P_{6} - 1)
$$
$$
+ \lambda_{10}( + P_{2} +  P_{4} - 1)+ \lambda_{11}( + P_{2} -  P_{4} - 1)+ \lambda_{12}( - P_{2} +  P_{4} - 1)+ \lambda_{13}( - P_{2} -  P_{4} - 1)
$$
$$
+ \lambda_{14}( + P_{2} +  P_{6} - 1)+ \lambda_{15}( + P_{2} -  P_{6} - 1)+ \lambda_{16}( - P_{2} +  P_{6} - 1)+ \lambda_{17}( - P_{2} -  P_{6} - 1)
$$
$$
+ \lambda_{18}( + P_{3} +  P_{4} - 1)+ \lambda_{19}( + P_{3} -  P_{4} - 1)+ \lambda_{20}( - P_{3} +  P_{4} - 1)+ \lambda_{21}( - P_{3} -  P_{4} - 1)
$$
$$
+ \lambda_{22}( + P_{3} +  P_{5} - 1)+ \lambda_{23}( + P_{3} -  P_{5} - 1)+ \lambda_{24}( - P_{3} +  P_{5} - 1)+ \lambda_{25}( - P_{5} -  P_{4} - 1)
$$
As noted in appendix A, any points that satisfy the KKT conditions are primal and dual optimal, and have zero duality gap so, we insist that points in FR must satisfy the KKT conditions which are
\begin{tabbing}
  1 \kill
  1. primal constraints:  \ \  $f_{i}\leqslant 0$ , i=1,...,25\\
  2. dual constraints:     \ \ $\lambda_{i}\geqslant0 , i=1,...,25$ \\
  3. complementary slackness:  \ \ $\lambda_{i} f_{i}(P_{1},...,P_{9})=0$ ,$i=1,...,25$ \\
  4. gradient of Lagrangian must vanish:\ \ $\nabla \mathbf{L}(P, \lambda, \nu)=0 .$
\end{tabbing}
\par
The first and second KKT conditions are satisfied automatically. For third constraints please note that duo to the convex optimization nature, the optimal point of the problem is in the intersection region of these constraints and, as this region also belong to the FR defined by only nonlinear constraint; therefore, we can consider the FR defined only by the nonlinear constraint and the role of other constraints are limiting this FR. Thus we can write the third condition of KKT in the following form
\begin{equation}\label{slack1}
    \lambda_{1}>0 \Rightarrow f_{1}(P_{1},...,P_{9})=0
\end{equation}
and
\begin{equation}\label{slack2}
    f_{1}(P_{1},...,P_{9}) < 0 \Rightarrow \lambda_{i}= 0, \quad i=2,...,25.
\end{equation}
Forth constraints of KKT conditions yields to
\begin{equation}\label{Pis}
    P_{i}=-\frac{A_{i}}{2 \lambda_{1}}, \quad i=1,...,9
\end{equation}
substituting these equations in (\ref{slack1}) gives
$$
4 \lambda_{1}^{2}=\sum_{i=1}^{9}A_{i}^{2}
$$
and the minimum value of $f(P_{1},...,P_{9})$ becomes
\begin{equation}\label{mincost}
    -( \ \sum_{i=1}^{9}A_{i}^{2} \ )^{\frac{1}{2}}
\end{equation}
\par
Now the EW takes form
\begin{equation}\label{wita}
    W=A_{0}III+\sum_{i=1}^{9}A_{i} Q_{i}
\end{equation}
with following constraint
\begin{equation}\label{conswita}
    A_{0}^{2} \geqslant \sum_{i=1}^{9}A_{i}^{2}
\end{equation}
This constraint ensure that $Tr(W \rho_{s})\geqslant 0$. The other constraints on EW (\ref{wita}) which comes from (\ref{const}), now takes form
$$
(A_{1}+A_{5})^{2}\leqslant R  \quad , (A_{1}-A_{5})^{2}\leqslant R
$$
$$
(A_{1}+A_{6})^{2}\leqslant R  \quad , (A_{1}-A_{6})^{2}\leqslant R
$$
$$
(A_{2}+A_{4})^{2}\leqslant R  \quad , (A_{2}-A_{4})^{2}\leqslant R
$$
$$
(A_{2}+A_{6})^{2}\leqslant R  \quad , (A_{2}-A_{6})^{2}\leqslant R
$$
$$
(A_{3}+A_{4})^{2}\leqslant R  \quad , (A_{3}-A_{4})^{2}\leqslant R
$$
\begin{equation}\label{newconst}
    (A_{3}+A_{5})^{2}\leqslant R  \quad , (A_{3}-A_{5})^{2}\leqslant R
\end{equation}
where
$R=\sum_{i=1}^{9} A_{i}^{2} .$
\par
For dual problem note that
$$
g(\lambda_{1})=-\lambda_{1}-\frac{1}{4 \lambda_{1}}\sum_{i=1}^{9}A_{i}^{2}
$$
so the dual problem take form
$$
 \mathrm{Maximize} \quad g(\lambda_{1})
$$
\begin{equation}\label{dualSphe}
    \mathrm{s.t.}  \quad \lambda_{1} > 0
\end{equation}
As $\lambda_{1} > 0$, the maximum value of $g(\lambda_{1})$ is
\begin{equation}\label{maxconst}
    -( \ \sum_{i=1}^{9}A_{i}^{2} \ )^{\frac{1}{2}}.
\end{equation}
So the minimum of primal problem (\ref{mincost}), is equal with the maximum of dual problem (\ref{maxconst}), and there is no gap between them and the minimum of primal problem is global.
\section{Constructing nonlinear EWs via convex optimization}
\par
As noted before, the envelope of family of linear EWs tangent to FR, could be considered as a nonlinear EW. We want to obtain this nonlinear EW via convex optimization. For this purpose we reformulate problem in convex optimization format. Suppose a density matrix, $\rho$, for a system is given. One can expand this density matrix in terms of related operators $Q_{i}$ in the Hilbert space of the system with coefficients say $r_{i}$.
$$
\rho=\sum_{i} r_{i} Q_{i}
$$
Entanglement detection condition requires $Tr(W \rho)\leqslant0$. Here $W$ is the family of linear EWs, which have been obtained from previous section. We want to minimize $Tr(W \rho)$ and the convex optimization problem takes the form
$$
\mathrm{minimize} \quad A_{0} r_{0}+ \sum_{i=1}^{n} A_{i} r_{i}
$$
\begin{equation}\label{w3}
    \mathrm{subject\quad to} \quad  f_{i}(A_{1},..,A_{n})\leqslant 0,
\end{equation}
where $f(A_{1},..,A_{n})$ is new inequality constraint which comes from previous section.
\par
So the nonlinear and linear EWs are constructed directly from convex optimization in two steps, as discussed above. This approach is completely general and could be applied for detection of entanglement of any quantum system. As a matter of fact, even for a system with complicating
nonlinear and linear constraints and functions, this approach will lead to some nonlinear and linear EWs, this is because of the convexity nature of the problem, and if there is no analytical solution to the problem, one can solve problem by good numerical algorithms such as interior point method (which again is valid for KKT conditions) \cite{Boyd}.
\par
In the previous works of obtaining nonlinear EWs with convex optimization \cite{JAAHM,JMHA}, there were two disadvantages. First, the linear inequality constraints are not considered, and second, the convex optimization for determining the nonlinear EWs, was not used explicitly in this form.

\emph{\textbf{Example 4: Nonlinear EW for three qubits system}} \label{exa2}
\par
In this example as a special case, we construct nonlinear EW for a given density matrix of three qubit system in example 3.
We choose EW in (\ref{wita}). In this step the linear inequality constraints takes form (\ref{newconst}). The given density matrix for three qubits system could be written as follows
$$
\rho=\sum_{i} r_{i} Q_{i}=\sum_{i,j,k=0}^{3}b_{i,j,k} \sigma_{i}\otimes\sigma_{j}\otimes\sigma_{k}
$$
For better detection of entanglement of the system, we want to minimize $Tr(W \rho)=\sum_{m=0}^{9}r_{m} A_{m}$ therefore, convex optimization problem takes form
\begin{equation}\label{secoconv}
    \mathrm{Minimize} \quad \sum_{m=0}^{9}r_{m} A_{m}
\end{equation}
$$
\mathrm{subject \quad to} \quad -A_{0}^{2}+R \leqslant0 ,
$$
$$
(A_{1}+A_{5})^{2}\leqslant R  \quad , (A_{1}-A_{5})^{2}\leqslant R
$$
$$
(A_{1}+A_{6})^{2}\leqslant R  \quad , (A_{1}-A_{6})^{2}\leqslant R
$$
$$
(A_{2}+A_{4})^{2}\leqslant R  \quad , (A_{2}-A_{4})^{2}\leqslant R
$$
$$
(A_{2}+A_{6})^{2}\leqslant R  \quad , (A_{2}-A_{6})^{2}\leqslant R
$$
$$
(A_{3}+A_{4})^{2}\leqslant R  \quad , (A_{3}-A_{4})^{2}\leqslant R
$$
$$
(A_{3}+A_{5})^{2}\leqslant R  \quad , (A_{3}-A_{5})^{2}\leqslant R
$$
The linear and nonlinear inequality constraints comes from (\ref{conswita}) and (\ref{newconst}).
The Lagrangian for this part of problem is
$$
\mathbf{L}(\mathbf{A}, \mathbf{\mu}) =A_{0} r_{0}+ \sum_{i=1}^{9} A_{i} r_{i} + \mu_{1}(\sum_{i=1}^{9}A_{i}^{2}-A_{0}^{^{2}})+\mu_{2}((A_{1}+A_{5})^{2}-R )+\mu_{3}((A_{1}-A_{5})^{2}-R )
$$
$$
+\mu_{4}((A_{1}+A_{6})^{2}-R )+\mu_{5}((A_{1}-A_{6})^{2}-R )
$$
$$
+\mu_{6}((A_{2}+A_{4})^{2}-R )+\mu_{7}((A_{2}-A_{4})^{2}-R )
$$
$$
+\mu_{8}((A_{2}+A_{6})^{2}-R )+\mu_{9}((A_{2}-A_{6})^{2}-R )
$$
$$
+\mu_{10}((A_{3}+A_{4})^{2}-R )+\mu_{11}((A_{3}-A_{4})^{2}-R )
$$
$$
+\mu_{12}((A_{3}+A_{5})^{2}-R )+\mu_{13}((A_{3}-A_{5})^{2}-R )
$$
The arguments for KKT conditions in example 3 are also valid here. From complementary slackness of KKT conditions we have
\begin{equation}\label{slack3}
    \mu_{1} > 0 \Rightarrow f_{1}(P_{1},...,P_{9})=\sum_{i=1}^{9}A_{i}^{2}-A_{0}^{^{2}}=0
\end{equation}
and
\begin{equation}\label{slack4}
    f_{i}(P_{1},...,P_{9})< 0 \Rightarrow \mu_{i}= 0 , \quad \mu=2,...,13
\end{equation}
and zero gradient of Lagrangian condition yields to
$$
A_{i}=-\frac{r_{i}}{2 \mu_{1}} , \quad i=1,...,9
$$
So the condition (\ref{slack3}) becomes
\begin{equation}\label{mucond}
    4 \mu_{1}^{2}=\frac{1}{A_{0}^{2}} \sum_{i=1}^{9}{r_{i}^{2}}
\end{equation}
the other constraints (\ref{slack4}), becomes
$$
(r_{1}+r_{5})^{2}\leqslant  T \quad , (r_{1}-r_{5})^{2}\leqslant  T
$$
$$
(r_{1}+r_{6})^{2}\leqslant  T \quad , (r_{1}-r_{6})^{2}\leqslant  T
$$
$$
(r_{2}+r_{4})^{2}\leqslant  T \quad , (r_{2}-r_{4})^{2}\leqslant  T
$$
$$
(r_{2}+r_{6})^{2}\leqslant  T \quad , (r_{2}-r_{6})^{2}\leqslant  T
$$
$$
(r_{3}+r_{4})^{2}\leqslant  T \quad , (r_{3}-r_{4})^{2}\leqslant  T
$$
\begin{equation}\label{lastcons}
    (r_{3}+r_{5})^{2}\leqslant  T \quad , (r_{3}-r_{5})^{2}\leqslant  T
\end{equation}
where
$$
T = \frac{R}{A_{0}^{2}}\sum_{i=1}^{9} r_{i}^{2}
$$
Thus the nonlinear EW detection becomes
\begin{equation}\label{du2}
    \mathrm{Min} \quad Tr(W \rho)=A_{0}[r_{0}-(\sum_{i=1}^{9} r_{i}^{2})^{\frac{1}{2}}]
\end{equation}
with constraints (\ref{lastcons}). For dual problem we have
$$
g(\mu_{1})=A_{0} r_{0}-(\mu_{1}+\frac{1}{4 \mu_{1}}\sum_{i=1}^{9}r_{i}^{2})
$$
so the dual problem take form
$$
 \mathrm{Maximize} \quad g(\mu_{1})
$$
\begin{equation}\label{dualSphe2}
    \mathrm{s.t.}  \quad \mu_{1} > 0
\end{equation}
As $\mu_{1} > 0$, the maximum value of $g(\mu_{1})$ is
\begin{equation}\label{maxconst2}
    A_{0}[r_{0}-(\sum_{i=1}^{9} r_{i}^{2})^{\frac{1}{2}}].
\end{equation}
Again, the minimum of primal problem (\ref{du2}), is equal with the maximum of dual problem (\ref{maxconst2}), and there is no gap between them and the minimum of primal problem is global.
\section{EWs for three qubits systems}
\par
There are many special sets of linear and nonlinear EWs for three qubits with specific FRs. In this section we recover some of them for three qubits systems. These are classified into four sets and finding these FRs and linear and nonlinear EWs are completely similar to the previous sections. In the following we report FRs and EWs concisely.
\subsection{EWs with polygonal FR}
\par
The polygonal FR in example 1, leads to polygonal class for three qubits linear EWs.
The convex optimization for this problem is
$$
\mathrm{minimize} \quad  A_{0}+\sum_{i=1}^{3} A_{i} P_{i}
$$
\begin{equation}\label{LinConvexOpti}
    \mathrm{subject\quad to} \quad \mathrm{equation} \quad (\ref{LiConst})
\end{equation}
and the relative EWs takes form
$$
 W_{_{i_{1},i_{2},i_{3},i_{4},i_{5}}}=A_{0}III+
$$
\begin{equation}\label{LinEWs}
   A_{1}[(-1)^{^{i_{1}}} Q_{1}+(-1)^{^{i_{2}}} Q_{2}+(-1)^{^{i_{3}}} Q_{3}+ (-1)^{^{i_{4}}}Q_{4}+(-1)^{^{i_{5}}} Q_{5}+(-1)^{^{i_{1}+i_{2}+i_{3}+i_{4}+i_{5}+1}} Q_{6}]
\end{equation}
where $i_{1},...,i_{5}=0,1$ then we have $32$ linear EWs. Besides these EWs, we can construct other EWs by using the fact that local unitary operators take an EW to another EW. The $36$ transformation of table (\ref{t1}) on (\ref{LinEWs}), give a new EW which could be constructed by local unitary operators. Please note that $M_{x\leftrightarrow y}^{2}$, means transformation which interchange $x$ and $y$ in the second qubit and so on. For example in (\ref{LinEWs}) for $i_{1}=...=i_{5}=0$ if we apply the transformation $M_{y\leftrightarrow z}^{1} M_{y\leftrightarrow z}^{2}$ then
$$
M_{y\leftrightarrow z}^{1} M_{y\leftrightarrow z}^{2} W_{_{0,0,0,0,0}}=A_{0}III+A_{1}[XXX+XZY+ZXZ+ZYY+YZZ-YYX]
$$
which is a new linear EW. Therefore the total linear EWs, with odd number minus signs and all transformations of table 1, becomes $32\times 36=1184$.
\par
These linear EW are non-decomposable because they can detect density matrices with positive partial transpose (PPT). For example the linear EW
$$
W=III+XXX+XYZ+YYY+YZX+ZXY-ZZZ
$$
which comes from applying transformation $M_{y\leftrightarrow z}^{3} M_{x\leftrightarrow y}^{2}M_{x\leftrightarrow y}^{3}$ on (\ref{LinEWs}) and taking $A_{0}=A_{1}=1, \quad i_{1}=...=i_{5}=0$; can detect the PPT density matrix in \cite{JAAHM}. As the non-decomposability of EWs are invariant under the transformations of table 1, therefore all $1184$ linear EWs in this section are also non-decomposable.
\par
\begin{table}
  \caption{\small{36 transformation on (\ref{LinEWs}).}\label{t1}}
  \begin{tabular}{|@{}l@{}|@{}l@{}|@{}l@{}|@{}l@{}|@{}l@{}|@{}l@{}}
    \hline
    \tiny{$M_{y\leftrightarrow z}^{3}$} &
    \tiny{$M_{x\leftrightarrow y}^{3}$} &
    \tiny{$M_{x\rightarrow y\rightarrow z\rightarrow x}^{3}$ }&
    \tiny{$M_{x\rightarrow z\rightarrow y\rightarrow x}^{3}$ }&
    \tiny{$M_{x\leftrightarrow z}^{3}$ }&
    \tiny{$M_{x\leftrightarrow y}^{2} M_{x\leftrightarrow y}^{3}$} \\ \hline
    \tiny{$M_{y\leftrightarrow z}^{2}$ }&
    \tiny{$M_{y\leftrightarrow z}^{1} M_{y\leftrightarrow z}^{2}$} &
    \tiny{$M_{y\leftrightarrow z}^{1} M_{x\rightarrow y\rightarrow z\rightarrow x}^{2}$} &
    \tiny{$M_{x\rightarrow y\rightarrow z\rightarrow x}^{2}$} &
    \tiny{$M_{x\rightarrow y\rightarrow z\rightarrow x}^{2}M_{x\leftrightarrow y}^{3}$ }&
    \tiny{$M_{x\leftrightarrow z}^{2}$ }\\ \hline
    \tiny{$M_{x\rightarrow y\rightarrow z\rightarrow x}^{1}$ }&
    \tiny{$M_{x\leftrightarrow z}^{1}$ }&
    \tiny{$M_{x\leftrightarrow z}^{2}M_{x\rightarrow y\rightarrow z\rightarrow x}^{3}$ }&
    \tiny{$M_{y\leftrightarrow z}^{3}M_{x\leftrightarrow y}^{2} M_{x\leftrightarrow y}^{3}$ }&
    \tiny{$M_{y\leftrightarrow z}^{3}M_{y\leftrightarrow z}^{2}$ }&
    \tiny{$M_{y\leftrightarrow z}^{3} M_{y\leftrightarrow z}^{1} M_{x\rightarrow y\rightarrow z\rightarrow x}^{2}$ } \\ \hline
    \tiny{$M_{x\leftrightarrow y}^{3}M_{x\leftrightarrow y}^{2} M_{x\leftrightarrow y}^{3}$ }&
    \tiny{$M_{x\rightarrow y\rightarrow z\rightarrow x}^{3}M_{x\leftrightarrow y}^{2} M_{x\leftrightarrow y}^{3}$ }&
    \tiny{$M_{x\leftrightarrow y}^{3}M_{y\leftrightarrow z}^{2}$ }&
    \tiny{$M_{x\leftrightarrow y}^{3} M_{y\leftrightarrow z}^{1}M_{y\leftrightarrow z}^{2} $ }&
    \tiny{$M_{x\rightarrow y\rightarrow z\rightarrow x}^{3}M_{y\leftrightarrow z}^{2}$ }&
    \tiny{$M_{x\rightarrow y\rightarrow z\rightarrow x}^{3}M_{y\leftrightarrow z}^{1}M_{y\leftrightarrow z}^{2}$ }\\ \hline
    \tiny{$M_{x\rightarrow y\rightarrow z\rightarrow x}^{3}M_{x\leftrightarrow y}^{2} M_{x\leftrightarrow y}^{3}$ }&
    \tiny{$M_{x\leftrightarrow z}^{3} $ }&
    \tiny{$M_{x\rightarrow z\rightarrow y\rightarrow x}^{3}M_{y\leftrightarrow z}^{2}$  }&
    \tiny{$M_{x\rightarrow z\rightarrow y\rightarrow x}^{3}M_{y\leftrightarrow z}^{1} M_{x\rightarrow y\rightarrow z\rightarrow x}^{2}$}&
    \tiny{$M_{x\leftrightarrow z}^{3}M_{y\leftrightarrow z}^{2}$ }&
    \tiny{$M_{x\leftrightarrow z}^{3}M_{y\leftrightarrow z}^{1} M_{x\rightarrow y\rightarrow z\rightarrow x}^{2}$ }\\ \hline
    \tiny{$M_{y\leftrightarrow z}^{3}M_{y\leftrightarrow z}^{1} M_{x\rightarrow y\rightarrow z\rightarrow x}^{2} $ }&
    \tiny{$M_{y\leftrightarrow z}^{3}M_{x\rightarrow y\rightarrow z\rightarrow x}^{2} $} &
    \tiny{$M_{x\leftrightarrow z}^{3}M_{x\leftrightarrow z}^{2} $ }&
    \tiny{$M_{x\leftrightarrow z}^{3}M_{x\rightarrow y\rightarrow z\rightarrow x}^{2}M_{x\leftrightarrow y}^{3} $ }&
    \tiny{$M_{x\leftrightarrow y}^{3}M_{x\rightarrow y\rightarrow z\rightarrow x}^{1} $ }&
    \tiny{ $M_{x\leftrightarrow y}^{3}M_{x\leftrightarrow z}^{1} $ }\\ \hline
  \end{tabular}
\end{table}
\subsection{EWs with conical FR}
\par
Let us consider the following operators
$$
Q_{1}^{Co}=Z(XX+YY),\quad Q_{2}^{Co}=X(XX+YY),\quad Q_{3}^{Co}=Y(XX+YY),
$$
$$
Q_{4}^{Co}=Z(XY-YX),\quad Q_{5}^{Co}=X(XY-YX),\quad Q_{6}^{Co}=Y(XY-YX),
$$
\begin{equation}\label{coco}
    Q_{13}^{Co}=IZZ.
\end{equation}
where the superscript $Co$ in $Q_{i}$'s, shows the conical case.
Now we try to determine the exact shape of the FR. The FR is a cone given by
\begin{equation}\label{con1}
    \sum_{i=1}^{6}P_{i}^{2}- (1\pm P_{13})^{2}\leqslant 0
\end{equation}
(for a proof, see appendix B).
First convex optimization gives two related EWs as follows
$$
\mathrm{minimize} \quad A_{13} P_{13}+ \sum_{i=1}^{6} A_{i} P_{i}
$$
\begin{equation}\label{w3}
    \mathrm{subject\quad to} \quad (\ref{con1})
\end{equation}
The minimum is equal to $-A_{13}$, provided that
\begin{equation}\label{cona}
    A_{13}^{2}=\sum_{i=1}^{6} A_{i}^{2}
\end{equation}
and the constraints $Tr(W\rho_{sep})\geqslant0$ leads to $A_{0}\geqslant A_{13}$. So the linear witnesses becomes
\begin{equation}\label{connon}
    W^{Co}=A_{0}(III\pm Q_{13}+\sum_{i=1}^{6}A_{i}Q_{i})
\end{equation}
\par
Second convex optimization gives the nonlinear EW as follows. The minimum of $Tr(W \rho)$ subject to constraints $A_{0}-A_{13}\geqslant0$ and (\ref{cona}) becomes
$$
Min \quad Tr(W \rho)=A_{0}(1\pm r_{13}-\sqrt{r_{1}^{2}+...+r_{6}^{2}}\quad )
$$
Further, if we consider other operators such as
$$
Q_{7}^{Co}=Z(XX-YY),\quad Q_{8}^{Co}=X(XX-YY),\quad Q_{9}^{Co}=Y(XX-YY),
$$
$$
Q_{10}^{Co}=Z(XY+YX),\quad Q_{11}^{Co}=X(XY+YX),\quad Q_{12}^{Co}=Y(XY+YX),
$$
then one can show that
$$
P_{1}^{2}+P_{4}^{2}=P_{7}^{2}+P_{10}^{2}
$$
$$
P_{2}^{2}+P_{5}^{2}=P_{8}^{2}+P_{11}^{2}
$$
\begin{equation}\label{sub}
    P_{3}^{2}+P_{6}^{2}=P_{9}^{2}+P_{12}^{2}
\end{equation}
One can get new EWs, under any replacement of one or more left hand sides of (\ref{sub}) with their respective right hand sides  in the (\ref{con1}). As this can be done in eight ways, number of EWs so far are $2\times8=16$. In addition, the replacement of first party with second or third also give new EWs and as a result, the number of EWs in this form become $16\times3=48$.
\par
Again similar to the previous subsection arguments, these EWs are non-decomposable.
\subsection{EWs with spherical FR}
\par
For some special choice of operators one can get FR with hyper spherical shape. Some set of these choices is for following operators.
$$
Q_{1}=Z(XX+YY), \quad Q_{2}=X(XX+YY), \quad Q_{3}=Y(XX+YY),
$$
$$
Q_{4}=Z(XY-YX), \quad Q_{5}=X(XY-YX), \quad Q_{6}=Y(XY-YX),
$$
$$
Q_{7}=Z(XX-YY), \quad Q_{8}=X(XX-YY), \quad Q_{9}=Y(XX-YY),
$$
$$
Q_{10}=Z(XY+YX), \quad Q_{11}=X(XY+YX), \quad Q_{12}=Y(XY+YX),
$$
\begin{equation}\label{Sph}
    Q_{13}=IXZ, \quad Q_{14}=IYZ, \quad Q_{15}=IZI.
\end{equation}
and the FR becomes
\begin{equation}\label{SphFR1}
    P_{1}^{2}+...+P_{6}^{2}+P_{13}^{2}+P_{14}^{2}+P_{15}^{2}\leqslant1
\end{equation}
The proof is similar to the previous proofs in the appendix B.
The relative EWs become
\begin{equation}\label{witSphCla}
    W=A_{0}III+A_{13} Q_{13}+A_{14} Q_{14}+A_{15} Q_{15}+\sum_{i=1}^{6}A_{i} Q_{i}
\end{equation}
with constraint $A_{0}^{2} \geqslant +A_{13}^{2}+A_{14}^{2}+A_{15}^{2}+\sum_{i=1}^{6}A_{i}^{2}$. Finding the nonlinear EWs is completely similar to example 4.
\par
We can find $14$ other FRs with the following replacements
$$
I\rightarrow (n_{1}X+n_{2}Y+n_{3}Z) ,\  n_{1}^{2}+n_{2}^{2}+n_{3}^{2}=1
$$
in any parties of $Q_{13}$, $Q_{14}$ or $Q_{15}$ in (\ref{Sph}). For example if we replace $Q_{13}=IXZ$ with $Q_{16}+Q_{17}+Q_{18}=XXZ+YXZ+ZXZ$, then the new FR becomes
\begin{equation}\label{sphFRRep}
    P_{1}^{2}+...+P_{6}^{2}+P_{14}^{2}+P_{15}^{2}+P_{16}^{2}+P_{17}^{2}+P_{18}^{2}\leqslant1
\end{equation}
Now the convex optimization problem is Min $\sum_{i=1}^{6}A_{i} P_{i}+\sum_{j=14}^{18}A_{j} P_{j}$,  s.t. $(\ref{sphFRRep})$.
and the linear EWs become
\begin{equation}\label{Sph22}
    W=A_{0}III+\sum_{i=1}^{6}A_{i} Q_{i}+\sum_{j=14}^{18}A_{j} Q_{j}
\end{equation}
with constraint $A_{0}\geqslant [\sum_{i=1}^{6}A_{i}^{2}+\sum_{j=14}^{18}A_{j}^{2}]^{1/2}$. Again this constraint comes from the condition $Tr(W\rho_{sep})\geqslant0$.
In each of these $1+14=15$ FRs if we replace one or more of left hand sides of the following equations with the respective right hand sides, we will get new FRs.
$$
P_{1}^{2}+P_{4}^{2}=P_{7}^{2}+P_{10}^{2}
$$
$$
P_{2}^{2}+P_{5}^{2}=P_{8}^{2}+P_{11}^{2}
$$
$$
P_{3}^{2}+P_{6}^{2}=P_{9}^{2}+P_{12}^{2}
$$
As there are $7$ possible replacements with one on replacement, we have $8\times15=120$ spherical FR up to now. In addition, the replacement of first operator with second or third in all terms of (\ref{Sph} ) also give new FR and as a result, we have $120\times 3=360$ spherical FR.
Finally learning from previous proofs in appendix B, we present another spherical FR which is
$$
\sum_{i=1}^{27}P_{i}^{2}\leqslant 1.
$$
Therefore the total EWs for this set becomes $361$ which are non-decomposable (similar to the previous subsections arguments).
\subsection{EWs with other FRs}
There are many FRs for three qubits which lead to relative EWs. Here we obtain two cases as follows.
\par
\emph{\textbf{A. First case}}
\par
Consider the following set of operators
$$
Q_{_{1}}=XZZ, \quad Q_{_{2}}=XXX, \quad Q_{_{3}}=ZXZ, \quad Q_{_{4}}=ZZX,
$$
$$
Q_{_{5}}=-ZXI, \quad Q_{_{6}}=-ZZI, \quad Q_{_{7}}=XZI, \quad Q_{_{8}}=XXI,
$$
$$
Q_{_{9}}=-IZX, \quad Q_{_{10}}=-IZZ, \quad Q_{_{11}}=IXZ, \quad Q_{_{12}}=IXX,
$$
\begin{equation}\label{upbset}
    Q_{_{13}}=-XIZ, \quad Q_{_{14}}=-ZIZ, \quad Q_{_{15}}=ZIX, \quad Q_{_{16}}=XIX.
\end{equation}
With this choice, we have the following nonlinear constraints ( the proof is similar to the previous proofs in appendix B, therefore is omitted ).
$$
(P_{_{1}}+P_{_{2}})^{^{2}}+(P_{3}+P_{4})^{2} \leqslant 1
$$
$$
P_{5}^{2}+P_{6}^{2}+P_{7}^{2}+P_{8}^{2} \leqslant 1
$$
$$
P_{9}^{2}+P_{10}^{2}+P_{11}^{2}+P_{12}^{2} \leqslant 1
$$
\begin{equation}\label{upbNI}
    P_{13}^{2}+P_{14}^{2}+P_{15}^{2}+P_{16}^{2} \leqslant 1
\end{equation}
The related EW is $W=A_{0}+\sum_{i=1}^{16}A_{i} Q_{i}$. As discussed before, using convex optimization method we see that this EW candid satisfy $Tr(W \rho_{s})\geqslant 0$ condition if
$$
A_{0}-(\sqrt{A_{1}^{2}+A_{3}^{2}}+ \sqrt{A_{5}^{2}+A_{6}^{2}+A_{7}^{2}+A_{8}^{2}}+
$$
\begin{equation}\label{upbEW1}
    \sqrt{A_{9}^{2}+A_{10}^{2}+A_{11}^{2}+A_{12}^{2}}+ \sqrt{A_{13}^{2}+A_{14}^{2}+A_{15}^{2}+A_{16}^{2}} ) \geqslant 0
\end{equation}
For obtaining nonlinear form now the inequality constraints have same form as in (\ref{upbNI}) but we must replace $P_{i}$ to $A_{i}$, so the convex optimization problem takes form
$$
\mathrm{Minimize} \quad Tr(W \rho)=A_{0}r_{0} + \sum_{j=1}^{16}A_{j}r_{j}
$$
subject to condition
$$
(A_{_{1}}+A_{_{2}})^{^{2}}+(A_{3}+A_{4})^{2} \leqslant 1
$$
$$
A_{5}^{2}+A_{6}^{2}+A_{7}^{2}+A_{8}^{2} \leqslant 1
$$
$$
A_{9}^{2}+A_{10}^{2}+A_{11}^{2}+A_{12}^{2} \leqslant 1
$$
\begin{equation}\label{upbNI2}
    A_{13}^{2}+A_{14}^{2}+A_{15}^{2}+A_{16}^{2} \leqslant 1
\end{equation}
Again using convex optimization method, the result is
$$
Tr(W\rho)=A_{0}[(r_{0}-(\sqrt{(r_{1}+r_{2})^{2}+(r_{3}+r_{4})^{2}}+ \sqrt{r_{5}^{2}+r_{6}^{2}+r_{7}^{2}+r_{8}^{2}}
$$
\begin{equation}\label{upbEWf}
    + \sqrt{r_{9}^{2}+r_{10}^{2}+r_{11}^{2}+r_{12}^{2}}+ \sqrt{r_{13}^{2}+r_{14}^{2}+r_{15}^{2}+r_{16}^{2}} )]
\end{equation}

\par
\emph{\textbf{B. Second case}}
\par
Here, we consider the following Hermitian operators
$$
Q_{1}=XXX,\quad  Q_{2}=YXX, \quad Q_{3}=ZXX, \quad Q_{4}=XYY,
$$
\begin{equation}\label{Roh5}
  \quad Q_{5}=YYY,\quad Q_{6}=ZYY, \quad Q_{7}=XZZ,\quad Q_{8}=YZZ,\quad  Q_{9}=ZZZ
\end{equation}
The FR takes form
\begin{equation}\label{5}
 \sqrt{P_{1}^{2}+P_{2}^{2}+P_{3}^{2}}+\sqrt{P_{4}^{2}+P_{5}^{2}+P_{6}^{2}}+\sqrt{P_{7}^{2}+P_{8}^{2}+P_{9}^{2}}\leqslant1.
\end{equation}
Using convex optimization method for satisfying the condition $Tr(W\rho)\geqslant0$ we have $A_{0}-\sqrt{A_{1}^{2}+A_{2}^{2}+A_{3}^{2}}\geqslant0$ with conditions $A_{1}^{2}+A_{2}^{2}+A_{3}^{2}=A_{4}^{2}+A_{5}^{2}+A_{6}^{2}=A_{7}^{2}+A_{8}^{2}+A_{9}^{2}.$
\\
The nonlinear EW takes form
\begin{equation}\label{sc}
    Tr(W\rho)=A_{0}{[1-(r_{1}^{2}+r_{2}^{2}+r_{3}^{2})^{1/2}+(r_{4}^{2}+r_{5}^{2}+r_{6}^{2})^{1/2}+(r_{7}^{2}+r_{8}^{2}+r_{9}^{2})^{1/2}]}
\end{equation}
\section{Optimality of the EWs }
\par
In general, we have not found a proof for optimality of nonlinear EWs although, we expect that optimality problem could be solved using convex optimization and this issue is currently under investigation. However, in this section we consider optimality proofs for linear EWs with polygonal FR, and a special case of spherical EWs.
\par
To do so let us recall that if there exist $\epsilon> 0$ and a
positive operator $ \mathcal{P}$ such that $W'=W-\epsilon
\mathcal{P}$ be again an EW, the EW $W$ is not optimal, otherwise
it is. Every positive operator can be expressed as a sum of pure
projection operators with positive coefficients, i.e.,
$\mathcal{P}=\sum_{i}\lambda_{_{i}}|\psi_{_{i}}\rangle\langle\psi_{_{i}}|$
with all $\lambda_{_{i}}\geq0$, so we can take $\mathcal{P}$  as
pure projection operator $\mathcal{P}=|\psi\rangle\langle\psi|$.
If $W'$ is to be an  EW, then $|\psi\rangle$ must be orthogonal to
all pure product states that the expectation value of W over them
is zero. The eigenstates of each three-qubit Pauli group operator
can be chosen as pure product states, half with eigenvalue +1 and
the other half with eigenvalue -1. In EWs introduced so far, there
exists no pair of locally commuting Pauli group operators, so the
expectation value of such pauli group operators vanishes over the
pure product eigenstates of one of them.
\subsection{\small{Optimality of the EWs with polygonal FR}}
\par
Let us begin with the following EWs with polygonal FR
$$
 W_{_{i_{1},i_{2},i_{3},i_{4},i_{5}}}=A_{0}III+
$$
\begin{equation}\label{optew}
   A_{1}[(-1)^{^{i_{1}}} ZZZ+(-1)^{^{i_{2}}} XXX+(-1)^{^{i_{3}}} XZY+ (-1)^{^{i_{4}}}YXZ+(-1)^{^{i_{5}}} YYY+(-1)^{^{i_{1}+i_{2}+i_{3}+i_{4}+i_{5}+1}}ZYX].
\end{equation}
This EW comes from  with transformation $M_{y\leftrightarrow z}^{2}$  on the(\ref{LinEWs}) and rearranging terms.
We discuss two cases $i_{_{1}}=0$ and $i_{_{1}}=1$ separately. For
the case $i_{_{1}}=0$, note that we can take the pure product
states
\begin{equation}\label{eigen1}
|z;+\rangle|z;+\rangle|z;+\rangle,\quad
|z;+\rangle|z;-\rangle|z;-\rangle,\quad
|z;-\rangle|z;+\rangle|z;-\rangle,\quad
|z;-\rangle|z;-\rangle|z;+\rangle,\quad
\end{equation}
as eigenstates of the operator $\sigma_{z}\sigma_{z}\sigma_{z}$
with eigenvalue +1 and the following ones
\begin{equation}\label{eigen2}
    |z;+\rangle|z;+\rangle|z;-\rangle,\quad
|z;+\rangle|z;-\rangle|z;+\rangle,\quad
|z;-\rangle|z;+\rangle|z;+\rangle,\quad
|z;-\rangle|z;-\rangle|z;-\rangle.
\end{equation}
as eigenstates with eigenvalue -1. The EWs
$W_{0,i_{_{2}},i_{_{3}},i_{_{4}},i_{_{5}}}$ have zero expectation values over the states of (\ref{eigen2}), so if there
exists a pure projection operator $|\psi\rangle\langle\psi|$ that
can be subtracted from EWs
$W_{0,i_{_{2}},i_{_{3}},i_{_{4}},i_{_{5}}}$, the state
$|\psi\rangle$ ought to be of the form
\begin{equation}\label{psi1}
\begin{array}{c}
  |\psi\rangle=a_{_{+++}}|z;+\rangle|z;+\rangle|z;+\rangle+a_{_{+--}}|z;+\rangle|z;-\rangle|z;-\rangle\\
 \hspace{1.3cm} +a_{_{-+-}}|z;-\rangle|z;+\rangle|z;-\rangle+a_{_{--+}}|z;-\rangle|z;-\rangle|z;+\rangle.\\
\end{array}
\end{equation}
Expectation values of $W_{0,0,i_{_{3}},i_{_{4}},i_{_{5}}}$
over pure product eigenstates of the operator
$\sigma_{x}\sigma_{x}\sigma_{x}$ with eigenvalue -1 are zero, so
$|\psi\rangle$ should be orthogonal to these eigenstates. Applying
the orthogonality constraints gives the following equations
$$
\begin{array}{c}
  \langle x;+|\langle x;+|\langle x;-||\psi\rangle=\frac{1}{2\sqrt{2}}(a_{_{+++}}-a_{_{+--}}-a_{_{-+-}}+a_{_{--+}})=0,\\
  \langle x;+|\langle x;-|\langle x;+||\psi\rangle=\frac{1}{2\sqrt{2}}(a_{_{+++}}-a_{_{+--}}+a_{_{-+-}}-a_{_{--+}})=0,\\
  \langle x;-|\langle x;+|\langle x;+||\psi\rangle=\frac{1}{2\sqrt{2}}(a_{_{+++}}+a_{_{+--}}-a_{_{-+-}}-a_{_{--+}})=0,\\
  \langle x;-|\langle x;-|\langle x;-||\psi\rangle=\frac{1}{2\sqrt{2}}(a_{_{+++}}+a_{_{+--}}+a_{_{-+-}}+a_{_{--+}})=0.\\
\end{array}
$$
The solution of this system of four linear equations is
$a_{_{+++}}=a_{_{+--}}=a_{_{-+-}}=a_{_{--+}}=0$. Thus
$|\psi\rangle=0$, that is, there exists no pure projection
operator $|\psi\rangle\langle\psi|$, hence no positive operator
$\mathcal{P}$, which can be subtracted from
$W_{0,0,i_{_{3}},i_{_{4}},i_{_{5}}}$ and leave them EWs again. So the EWs $W_{0,0,i_{_{3}},i_{_{4}},i_{_{5}}}$ are
optimal. Similar argument proves the optimality of EWs
$W_{0,1,i_{_{3}},i_{_{4}},i_{_{5}}}$.
\par
As for EWs $W_{1,i_{_{2}},i_{_{3}},i_{_{4}},i_{_{5}}}$, the state $|\psi\rangle$ (if exises) ought to be of the form
\begin{equation}\label{psi2}
\begin{array}{c}
  |\psi\rangle=a_{_{++-}}|z;+\rangle|z;+\rangle|z;-\rangle+a_{_{+-+}}|z;+\rangle|z;-\rangle|z;+\rangle\\
 \hspace{1.3cm} +a_{_{-++}}|z;-\rangle|z;+\rangle|z;+\rangle+a_{_{---}}|z;-\rangle|z;-\rangle|z;-\rangle.\\
\end{array}
\end{equation}
The same argument as above shows the impossibility of existing
such $|\psi\rangle$. Therefore, the EWs
$W_{1,i_{_{2}},i_{_{3}},i_{_{4}},i_{_{5}}}$ are also optimal.
\subsection{\small{Optimality of a special case of spherical EWs}}
For some special cases of EWs with spherical FR one can show the optimality of EWs. For example, consider the following case which is the spherical case in \cite{JAAHM}.
\par
One of these spherical EWs is
\begin{equation}\label{jaahmew}
    W=III+\frac{1}{\sqrt{A_{1}^{2}+A_{2}^{2}+A_{3}^{2}}}[A_{1}ZII+A_{2}(XXX+XYY)+A_{3}(YXY+YYX)]
\end{equation}
Let us first find pure product states that the expectation value of
(\ref{jaahmew}) over them vanishes. For this
purpose, we consider a pure product state as follows
\begin{equation}\label{pupr}
|\nu\rangle=\bigotimes_{j=1}^{3}\left(\cos(\frac{\theta_{_{j}}}{2})
|z;+\rangle+\exp(i\varphi_{_{j}})\sin(\frac{\theta_{_{j}}}{2})|z;-\rangle\right)
\end{equation}
and attempt to choose parameters $\theta_{_{j}}$ and
$\varphi_{_{j}}$ such that
$Tr(W|\nu\rangle\langle\nu|)=0$.
By direct calculation, this trace is
\begin{equation}\label{trace1}
\begin{array}{c}
 \hspace{-5cm} Tr(W|\nu\rangle\langle\nu|)=1+\frac{A_{1}}{{\sqrt{A_{1}^{2}+A_{2}^{2}+A_{3}^{2}}}}\cos\theta_{_{1}} +\sin\theta_{_{1}}\sin\theta_{_{2}}\sin\theta_{_{3}}\\
  \times[\frac{A_{2}}{{\sqrt{A_{1}^{2}+A_{2}^{2}+A_{3}^{2}}}}\cos\varphi_{_{1}}\cos(\varphi_{_{2}}-\varphi_{_{3}})+
  \frac{A_{3}}{{\sqrt{A_{1}^{2}+A_{2}^{2}+A_{3}^{2}}}}\sin\varphi_{_{1}}\sin(\varphi_{_{2}}+\varphi_{_{3}})]. \\
\end{array}
\end{equation}
In this relation, if we choose $\varphi_{_{2}}=\varphi_{_{3}}=\frac{\pi}{4}$, $\cos\psi_{_{1}}=\frac{A_{2}}{{\sqrt{A_{2}^{2}+A_{3}^{2}}}}$, and $\sin\psi_{_{1}}=\frac{A_{3}}{{\sqrt{A_{2}^{2}+A_{3}^{2}}}}$ then (\ref{trace1}) will become
\begin{equation}\label{trace2}
\begin{array}{c}
 Tr(W|\nu\rangle\langle\nu|)=1+\frac{A_{1}}{{\sqrt{A_{1}^{2}+A_{2}^{2}+A_{3}^{2}}}}\cos\theta_{_{1}} +\sin\theta_{_{1}}\sin\theta_{_{2}}\sin\theta_{_{3}}
  \frac{A_{2}}{{\sqrt{A_{1}^{2}+A_{2}^{2}+A_{3}^{2}}}}\cos(\psi_{_{1}}-\varphi_{_{1}}) \\
\end{array}
\end{equation}
In (\ref{trace1}), if we choose $\psi_{_{1}}=\varphi_{_{1}}$, $\theta_{_{2}}=\theta_{_{3}}=\frac{\pi}{2}$, $\cos\psi_{_{2}}=\frac{A_{1}}{{\sqrt{A_{1}^{2}+A_{2}^{2}+A_{3}^{2}}}}$, and $\sin\psi_{_{2}}=\frac{\sqrt{A_{2}^{2}+A_{3}^{2}}}{{\sqrt{A_{1}^{2}+A_{2}^{2}+A_{3}^{2}}}}$ then (\ref{trace2}) will become
\begin{equation}\label{trace2}
    Tr(W|\nu\rangle\langle\nu|)=1+\cos(\psi_{_{2}}-\theta_{_{1}}).
\end{equation}
and the choices of parameters $\psi_{_{2}}-\theta_{_{1}}=\pi$, lead to zero value for the
$Tr(W|\nu\rangle\langle\nu|)=0$.
\par
Now similar to the above discussion, it is easy to see that the following eight choices of parameters
$\theta_{_{j}}$ and $\varphi_{_{j}}$ lead to zero value for the $Tr(W|\nu\rangle\langle\nu|)$ :
$$
\begin{array}{c}
  \hspace{-.4cm} |\nu_{_{1}}\rangle:\quad
\theta_{_{2}}=\theta_{_{3}}=\frac{\pi}{2},\quad \psi_{_{2}}-\theta_{_{1}}=\pi
  ,\quad \varphi_{_{1}}=\psi_{_{1}},\quad \varphi_{_{2}}=\varphi_{_{3}}=\frac{\pi}{4},\\
  \hspace{-.4cm} |\nu_{_{2}}\rangle:\quad
\theta_{_{2}}=\theta_{_{3}}=\frac{\pi}{2},\quad \psi_{_{2}}-\theta_{_{1}}=\pi
  ,\quad \varphi_{_{1}}=-\psi_{_{1}},\quad \varphi_{_{2}}=\varphi_{_{3}}=-\frac{\pi}{4},\\
  \hspace{-.4cm} |\nu_{_{3}}\rangle:\quad
\theta_{_{2}}=\theta_{_{3}}=\frac{\pi}{2},\quad \psi_{_{2}}+\theta_{_{1}}=\pi
  ,\quad \varphi_{_{1}}=\psi_{_{1}},\quad \varphi_{_{2}}=\frac{\pi}{4},\quad \varphi_{_{3}}=-\frac{3\pi}{4},\\
  \hspace{-.4cm} |\nu_{_{4}}\rangle:\quad
\theta_{_{2}}=\theta_{_{3}}=\frac{\pi}{2},\quad \psi_{_{2}}+\theta_{_{1}}=\pi
  ,\quad \varphi_{_{1}}=-\psi_{_{1}},\quad \varphi_{_{2}}=\frac{3\pi}{4},\quad \varphi_{_{3}}=-\frac{\pi}{4},\\
  \hspace{-.4cm} |\nu_{_{5}}\rangle:\quad
\theta_{_{2}}=\theta_{_{3}}=\frac{\pi}{2},\quad \psi_{_{2}}-\theta_{_{1}}=\pi
  ,\quad \varphi_{_{1}}=\psi_{_{1}},\quad \varphi_{_{2}}=\frac{5\pi}{4},\quad \varphi_{_{3}}=-\frac{3\pi}{4},\\
  \hspace{-.4cm} |\nu_{_{6}}\rangle:\quad
\theta_{_{2}}=\theta_{_{3}}=\frac{\pi}{2},\quad \psi_{_{2}}-\theta_{_{1}}=\pi
  ,\quad \varphi_{_{1}}=-\psi_{_{1}},\quad \varphi_{_{2}}=\frac{3\pi}{4},\quad \varphi_{_{3}}=-\frac{5\pi}{4},\\
  \hspace{-.4cm} |\nu_{_{7}}\rangle:\quad
\theta_{_{2}}=\theta_{_{3}}=\frac{\pi}{2},\quad \psi_{_{2}}+\theta_{_{1}}=\pi
  ,\quad \varphi_{_{1}}=\psi_{_{1}},\quad \varphi_{_{2}}=-\frac{3\pi}{4},\quad \varphi_{_{3}}=\frac{\pi}{4},\\
  \hspace{-.4cm} |\nu_{_{8}}\rangle:\quad
\theta_{_{2}}=\theta_{_{3}}=\frac{\pi}{2},\quad \psi_{_{2}}+\theta_{_{1}}=\pi
  ,\quad \varphi_{_{1}}=-\psi_{_{1}},\quad \varphi_{_{2}}=-\frac{\pi}{4},\quad \varphi_{_{3}}=\frac{3\pi}{4}.\\
\end{array}
$$
For (\ref{jaahmew}), the state $|\psi\rangle$ (if exists) must be of the following form
\begin{equation}\label{psi22}
\begin{array}{c}
  |\psi\rangle=a_{_{+++}}|z;+\rangle|z;+\rangle|z;+\rangle+a_{_{++-}}|z;+\rangle|z;+\rangle|z;-\rangle\\
 \hspace{1.3cm} +a_{_{+-+}}|z;+\rangle|z;-\rangle|z;+\rangle+a_{_{+--}}|z;+\rangle|z;-\rangle|z;-\rangle\\
   +a_{_{-++}}|z;-\rangle|z;+\rangle|z;+\rangle+a_{_{-+-}}|z;-\rangle|z;+\rangle|z;-\rangle\\
 \hspace{1.3cm} +a_{_{--+}}|z;-\rangle|z;-\rangle|z;+\rangle+a_{_{---}}|z;-\rangle|z;-\rangle|z;-\rangle.\\
\end{array}
\end{equation}
and be orthogonal to the above eight states, i.e.,
$$
\begin{array}{c}
  \langle\nu_{_{1}}|\psi\rangle=
  [\sin\frac{\psi_{2}}{2}(a_{_{+++}} +a_{_{++-}}e^{i\frac{\pi}{4}}+a_{_{+-+}}e^{i\frac{\pi}{4}}+a_{_{+--}}e^{i\frac{\pi}{2}})\\
  -e^{i\psi_{1}}\cos\frac{\psi_{2}}{2}(a_{_{-++}}+a_{_{-+-}}e^{i\frac{\pi}{4}}+a_{_{--+}}e^{i\frac{\pi}{4}}
  +a_{_{---}}e^{i\frac{\pi}{2}})]=0,\\
  \langle\nu_{_{2}}|\psi\rangle=
  [\sin\frac{\psi_{2}}{2}(a_{_{+++}} +a_{_{++-}}e^{i\frac{-\pi}{4}}+a_{_{+-+}}e^{i\frac{-\pi}{4}}+a_{_{+--}}e^{i\frac{-\pi}{2}})\\
  -e^{-i\psi_{1}}\cos\frac{\psi_{2}}{2}(a_{_{-++}}+a_{_{-+-}}e^{i\frac{-\pi}{4}}+a_{_{--+}}e^{i\frac{-\pi}{4}}
  +a_{_{---}}e^{i\frac{-\pi}{2}})]=0,\\
  \langle\nu_{_{3}}|\psi\rangle=
  [\sin\frac{\psi_{2}}{2}(a_{_{+++}} -a_{_{++-}}e^{i\frac{\pi}{4}}+a_{_{+-+}}e^{i\frac{\pi}{4}}-a_{_{+--}}e^{i\frac{\pi}{2}})\\
  +e^{i\psi_{1}}\cos\frac{\psi_{2}}{2}(a_{_{-++}}-a_{_{-+-}}e^{i\frac{\pi}{4}}+a_{_{--+}}e^{i\frac{\pi}{4}}
  -a_{_{---}}e^{i\frac{\pi}{2}})]=0,\\
  \langle\nu_{_{4}}|\psi\rangle=
  [\sin\frac{\psi_{2}}{2}(a_{_{+++}} +a_{_{++-}}e^{i\frac{-\pi}{4}}-a_{_{+-+}}e^{i\frac{-\pi}{4}}-a_{_{+--}}e^{-i\frac{\pi}{2}})\\
  +e^{-i\psi_{1}}\cos\frac{\psi_{2}}{2}(a_{_{-++}}+a_{_{-+-}}e^{i\frac{-\pi}{4}}-a_{_{--+}}e^{i\frac{-\pi}{4}}
  -a_{_{---}}e^{-i\frac{\pi}{2}})]=0,\\
  \langle\nu_{_{5}}|\psi\rangle=
  [\sin\frac{\psi_{2}}{2}(a_{_{+++}} -a_{_{++-}}e^{i\frac{\pi}{4}}-a_{_{+-+}}e^{i\frac{\pi}{4}}+a_{_{+--}}e^{i\frac{\pi}{2}})\\
  +e^{i\psi_{1}}\cos\frac{\psi_{2}}{2}(-a_{_{-++}}+a_{_{-+-}}e^{i\frac{\pi}{4}}+a_{_{--+}}e^{i\frac{\pi}{4}}
  -a_{_{---}}e^{i\frac{\pi}{2}})]=0,\\
  \langle\nu_{_{6}}|\psi\rangle=
  [\sin\frac{\psi_{2}}{2}(a_{_{+++}} -a_{_{++-}}e^{i\frac{-\pi}{4}}-a_{_{+-+}}e^{i\frac{-\pi}{4}}+a_{_{+--}}e^{i\frac{-\pi}{2}})\\
  +e^{-i\psi_{1}}\cos\frac{\psi_{2}}{2}(-a_{_{-++}}+a_{_{-+-}}e^{i\frac{-\pi}{4}}+a_{_{--+}}e^{-i\frac{\pi}{4}}
  -a_{_{---}}e^{i\frac{-\pi}{2}})]=0,\\
  \langle\nu_{_{7}}|\psi\rangle=
  [\sin\frac{\psi_{2}}{2}(a_{_{+++}} +a_{_{++-}}e^{i\frac{\pi}{4}}-a_{_{+-+}}e^{i\frac{\pi}{4}}-a_{_{+--}}e^{i\frac{\pi}{2}})\\
  +e^{i\psi_{1}}\cos\frac{\psi_{2}}{2}(a_{_{-++}}+a_{_{-+-}}e^{i\frac{\pi}{4}}-a_{_{--+}}e^{i\frac{\pi}{4}}
  -a_{_{---}}e^{i\frac{\pi}{2}})]=0,\\
   \langle\nu_{_{8}}|\psi\rangle=
  [\sin\frac{\psi_{2}}{2}(a_{_{+++}} -a_{_{++-}}e^{-i\frac{\pi}{4}}+a_{_{+-+}}e^{-i\frac{\pi}{4}}-a_{_{+--}}e^{-i\frac{\pi}{2}})\\
  +e^{-i\psi_{1}}\cos\frac{\psi_{2}}{2}(a_{_{-++}}-a_{_{-+-}}e^{-i\frac{\pi}{4}}+a_{_{--+}}e^{-i\frac{\pi}{4}}
  -a_{_{---}}e^{-i\frac{\pi}{2}})]=0,\\
\end{array}
$$
The above system of eight equations has trivial solution
$a_{_{+++}}=a_{_{++-}}=a_{_{+-+}}=a_{_{+--}}=a_{_{-++}}=a_{_{-+-}}=a_{_{--+}}=a_{_{---}}=0$ provided that
$\psi_{1}\neq 0, \pm\frac{\pi}{2},\pm \pi$ and $\psi_{2}\neq 0, \pm \pi$. This proves the
optimality of (\ref{jaahmew}) for all but $\psi_{1}= 0, \pm\frac{\pi}{2},\pm \pi$ and $\psi_{2}= 0, \pm \pi$ values of $\psi$.

\section{Detection of entanglement for three qubits systems}
\par
In this section we develop two applications for EWs obtained via convex optimization method. Firstly, a density matrix is given and we want to construct some EWs for determining entanglement of this density matrix, and secondly a general class of nonlinear EWs is known and we would like to find some density matrices which could be detected efficiently by this class of nonlinear EWs. First application is completely natural and some straightforward. But second is not so trivial and an important question is: what is the physical motivation for this constructed density matrix? Some motivation are as follows. As any density matrix shows a real physical system, the entanglement source and channels may be rearrange in a way that the final density matrix for system be equal approximately to the constructed density matrix. Although this is a hard task, but if the constructed density matrix is valuable from experimental point of view, maybe this procedure will become a way for entanglement detection. On the other hand, at least as a toy model, this method will give some intuition to physical system. Although we are not deal to these subjects, we will discuss about how to construct some density matrices by this method.
\subsection{Detection of density matrices}
\par
We begin with some known density matrices for three qubits systems and try to detect entanglement of them with nonlinear EW constructed by exact convex optimization in the following three examples.
\par
\emph{\textbf{A. Unextendible product bases density matrix}}
\par
The density matrix considered here, is the entangled state in \cite{UPBD} which is constructed using unextendible product bases (UPBs), and has the very interesting property of being separable for every possible bipartition of the three parties. The state has the following expression:
\begin{equation}\label{upb}
    \rho=\frac{1}{4}(III-\sum_{i=1}^{4}|\psi_{i}\rangle \langle \psi_{i}|),
\end{equation}
where
$$
\psi_{1}=|0,1,+\rangle, \psi_{2}=|1,+,0\rangle,
$$
$$
\psi_{3}=|+,0,1\rangle, \psi_{4}=|-,-,-\rangle,
$$
and $|\pm\rangle=(|0\rangle\pm |1\rangle)/\sqrt{2}.$
Rewriting (\ref{upb}) in terms of Pauli operators yields
$$
\rho=\frac{1}{8}[III +\frac{1}{4}(-IXX-IXZ+IZX+IZZ-XIX+XIZ
$$
\begin{equation}\label{upbPauli}
    -XXI+XXX-XZI+XZZ-ZIX+ZIZ+ZXI+ZXZ+ZZI+ZZX) ]
\end{equation}
Now one can choose operators sets $Q_{i}$ from (\ref{upbPauli}) in a way that the related EW can detect entanglement of the system. One of these sets are (\ref{upbset}) which leads to the nonlinear EW (\ref{upbEWf}). The detection of this nonlinear EW (\ref{upbEWf}) for UPB density matrix (\ref{upbPauli}) is $Tr(W \rho)=\frac{-1-\sqrt{2}}{16} $.
\par
\emph{\textbf{B. W state density matrices}}
\par
The second mixed state density matrix which we consider here is W state density matrix \cite{Acin}. Consider the state
$$
\rho=\frac{1}{8}(1-p)III+p |W\rangle\langle W|
$$
where $|W\rangle=|100\rangle+|010\rangle+|001\rangle$ is the three partite W state. In \cite{Acin}, using an entanglement witness operator, the range for the parameter p, in which their EW detects $\rho$, i.e., $Tr(W \rho)<0$, is found to be $3/5 < p \leqslant 1$. \par
Using our nonlinear EW (\ref{sc}), the entanglement detection range for parameter p, is $3/7<p\leqslant 1$ which shows better detection (range of p is wider than before).
\par
\emph{\textbf{C. Mixed GHZ with W states density matrices}}
\par
As the final example consider the following mixed GHZ with W states density matrix
$$
\rho=\frac{1}{4}|\psi_{1}\rangle\langle \psi_{1}|+\frac{3}{8}(|W_{1}\rangle\langle W_{1}|+|W_{2}\rangle\langle W_{2}|)
$$
where $|\psi_{1}\rangle=\frac{1}{\sqrt{2}}(|000\rangle\pm|111\rangle)$ is GHZ state for three-qubits and $|W_{1}\rangle=\frac{1}{\sqrt{3}}(|001\rangle+|010\rangle+|100\rangle)$, $|W_{2}\rangle=\frac{1}{\sqrt{3}}(|110\rangle+|101\rangle+|011\rangle)$ are W states for three-qubits. The nonlinear EW (\ref{upbEWf}), can detect the entanglement of this density matrix and the detection is $Tr(W\rho)=-\frac{3}{32}$.

\subsection{Miscellaneous three-qubits PPT density matrices}
Here we construct some  three-qubits PPT density matrices by nonlinear EWs. As an example consider the following nonlinear EW.
\begin{equation}\label{condenmat}
    W=A_{0}(III\pm Q_{13}^{Co} + \sum_{i=1}^{3}A_{i} Q_{i}^{Co}+ \sum_{j=10}^{12}A_{j} Q_{j}^{Co})
\end{equation}
Now we choose some Pauli operators from this nonlinear EW and introduce a density matrix in the following form
$$
\rho=\frac{1}{8}[III+r_{1}IZZ+r_{2}(ZXX+ZYY)+r_{3}(XXX+XYY)+r_{4}(YXX+YYY)
$$
\begin{equation}\label{2kindConical}
   +r_{5}(ZXY+ZYX)+r_{6}(XXY+XYX)+r_{7}(YXY+YYX)]
\end{equation}
The PPT conditions for this density matrix are
\begin{equation}\label{ppt1}
    ( 1\pm r_{1} \pm 2 \sqrt{R_{1}^{2}}\ )\geqslant 0,
\end{equation}
\begin{equation}\label{ppt1}
    ( 1\pm r_{1} \pm 2 \sqrt{R_{2}^{2}}\ )\geqslant 0.
\end{equation}
where $R_{1}^{2}= r_{2}^{2}+r_{3}^{2}+r_{4}^{2}$ and $R_{2}^{2}=r_{5}^{2}+r_{6}^{2}+r_{7}^{2}$. The solution for these PPT conditions are
\begin{equation}\label{sol1}
    r_{1}=1, \quad R_{1}=0, \quad R_{2}=0
\end{equation}
\begin{equation}\label{sol2}
    r_{1}=-1, \quad R_{1}=0, \quad R_{2}=0
\end{equation}
\begin{equation}\label{sol3}
    -1< r_{1}\leqslant 0, \quad -(1+r_{1})\leqslant 2R_{1}\leqslant(1+r_{1}), \quad -(1+r_{1})\leqslant 2R_{2}\leqslant(1+r_{1})
\end{equation}
\begin{equation}\label{sol4}
     0< r_{1}<1, \quad -(1-r_{1})\leqslant 2R_{1}\leqslant(1-r_{1}), \quad -(1-r_{1})\leqslant 2R_{2}\leqslant(1-r_{1})
\end{equation}
so the detection conditions become
$$
Tr(W \rho)= 1+r_{1}-2 \sqrt{R_{1}^{2}+R_{2}^{2}} < 0,
$$
$$
Tr(W \rho)= 1-r_{1}-2 \sqrt{R_{1}^{2}+R_{2}^{2}} < 0.
$$
Therefore we construct a three qubits PPT density matrix by a nonlinear EW and the entanglement of the density matrix is detected by this nonlinear EW.
\par
As another example, consider following density matrix
\begin{equation}\label{Rohl}
   \rho=\frac{1}{8}[III+r_{1}XXX+r_{2}YXX+r_{3}ZXX+r_{4}XYY+r_{5}YYY+r_{6}ZYY+r_{7}XZZ+r_{8}YZZ+r_{9}ZZZ]
\end{equation}
The PPT conditions for this density matrix are
\begin{equation}\label{Roh2}
    \frac{1}{8}(1\pm\sqrt{(r_{1}+r_{4}-r_{7})^{2}+(r_{2}+r_{5}-r_{8})^{2}+(r_{3}+r_{6}-r_{9})^{2}})\geqslant0
\end{equation}
\begin{equation}\label{Roh3}
    \frac{1}{8}(1\pm\sqrt{(r_{1}-r_{4}+r_{7})^{2}+(r_{2}-r_{5}+r_{8})^{2}+(r_{3}-r_{6}+r_{9})^{2}})\geqslant0
\end{equation}
\begin{equation}\label{Roh4}
   \frac{1}{8}(1\pm\sqrt{(-r_{1}+r_{4}+r_{7})^{2}+(-r_{2}+r_{5}+r_{8})^{2}+(-r_{3}+r_{6}+r_{9})^{2}})\geqslant0
\end{equation}
\begin{equation}\label{Roh5}
   \frac{1}{8}(1\pm\sqrt{(r_{1}+r_{4}+r_{7})^{2}+(r_{2}+r_{5}+r_{8})^{2}+(r_{3}+r_{6}+r_{9})^{2}})\geqslant0
\end{equation}
For this case, the non-linear EW, (\ref{sc}), detects (\ref{Rohl}) with following conditions.
In the PPT conditions if we choose $r_{i} \geqslant0$ for $\forall i$ and also the final PPT condition (\ref{Roh5}) is satisfied, then all other PPT conditions would be satisfied. In addition, the following three inequalities must be satisfied (for detection)
$$
[(r_{1}^{2}+r_{2}^{2}+r_{3}^{2})(r_{4}^{2}+r_{5}^{2}+r_{6}^{2})]^{1/2} \geqslant (r_{1}r_{4}+r_{2}r_{5}+r_{3}r_{6}),
$$
$$
[(r_{1}^{2}+r_{2}^{2}+r_{3}^{2})(r_{7}^{2}+r_{8}^{2}+r_{9}^{2})]^{1/2} \geqslant (r_{1}r_{7}+r_{2}r_{8}+r_{3}r_{9}),
$$
$$
[(r_{4}^{2}+r_{5}^{2}+r_{6}^{2})(r_{7}^{2}+r_{8}^{2}+r_{9}^{2})]^{1/2} \geqslant (r_{5}r_{7}+r_{5}r_{8}+r_{6}r_{9}).
$$
which come from applying the Cauchy-Schwartz inequality to each part.
\section{Conclusion}
\par
We have presented a general algorithm via exact convex optimization to the problem of finding nonlinear and linear EWs.
This approach is completely general and could be applied for detection of entanglement of any N-partite quantum system. For this purpose we defined a map from convex space of separable density matrices to a convex region called FR so the problem of finding EWs was reduced to the convex optimization problem which could be solved by the Karush-Kuhn-Tucker convex optimization method. The problem of finding FRs is occupy a significant place in our algorithm and the main finding of the study for three-qubits reveal how systematic such convex optimization algorithm can be. As exemplified by our three-qubits study, there are many FRs for a quantum system which lead to linear and nonlinear EWs and this is a good reason to think that finding the whole FRs is time-consuming and our expectation is that finding the whole FR is a nontrivial algebraic geometry problem. While our analysis is for three-qubits systems, it serves to provide a unified explanation for a variety of EWs with striking detection ability with respect to previous EWs.
The main conclusion is that the presented algorithm provide indispensable prerequisites for further investigation and can bring a
robustness in constructing EWs for a system. Application of this algorithm to other quantum system and finding related FR is still an open problem which is under investigation.

\newpage
\vspace{1cm}\setcounter{section}{0}
\setcounter{equation}{0}
\renewcommand{\theequation}{A-\roman{equation}}
  {\Large{Appendix A}}\\
\textbf{Convex optimization review }
\par
An optimization problem  \cite{Boyd}, has the standard form\\
minimize \ \ $f_{0}(x)$\\
subject to \ $f_{i}(x)\leqslant 0$,   $i=1,...,m.$ \\
             $h_{i}(x)= 0$,   $i=1,...,p.$ \\
Where the vector $x=(x_{1},...,x_{n})$ is the optimization variable of the problem, the function $f_{0} : \textbf{R}^{n} \rightarrow \textbf{R}$ is the objective function, the functions $f_{i} : \textbf{R}^{n} \rightarrow \textbf{R}$, $i=1,...,m$, are the (inequality) constraint functions, and the constants $b_{1},...,b_{m}$ are the limits, or bounds, for the constraints.
A convex optimization problem, is an optimization problem where the objective and the constraint functions are convex functions which means they satisfy inequality $f_{i}(\alpha x +\beta y)\leqslant \alpha f_{i}(x)+\beta f_{i}(y)$,
for all $x, y \in R$ and all $\alpha, \beta \in R$ with $\alpha +\beta = 1$, $\alpha \geqslant 0$, $\beta \geqslant 0$ and the equality constraint functions $h_{i}(x)= 0$ must be affine (A set $C\in\textbf{R}^{n}$ is affine if the line through any two distinct points in $C$ lies in $C$).
\par
One can solve this convex optimization problem using Lagrangian duality. The basic idea in the Lagrangian duality is to take the constraints in convex optimization problem into account by augmenting the objective function with a weighted sum of the constraint
functions. The Lagrangian L : $\textbf{R}^{n}\times \textbf{R}^{m} \times\textbf{R}^{p}\rightarrow \textbf{R}$  associated with the
problem is defined as
\begin{equation}\label{lag}
    L(x, \lambda, \nu)=f_{0}(x) + \sum_{i=1}^{m}\lambda_{i} f_{i}(x)+\sum_{i=1}^{p}\nu_{i} h_{i}(x)
\end{equation}
The Lagrange dual function $g: \textbf{R}^{m} \times\textbf{R}^{n}\times \textbf{R}^{m} \times\textbf{R}^{p}\rightarrow \textbf{R}$ is defined as the minimum value of the Lagrangian over x: for $\lambda \in \textbf{R}^{m}, \nu \in \textbf{R}^{p}$,
\begin{equation}\label{duf}
    g(\lambda, \nu)=inf_{_{x\in \textbf{D}}} L(x, \lambda, \nu)
\end{equation}
The dual function yields lower bounds on the optimal value $p^{\star}$ of the convex optimization problem, i.e for any $\lambda\succeq0$ and any $\nu$ we have
\begin{equation}\label{lowerBound}
    g(\lambda, \nu)\leqslant p^{\star}
\end{equation}
\par
The optimal value of the Lagrange dual problem, which we denote $d^{\star}$, is, by definition, the best lower bound on $d^{\star}$ that can be obtained from the Lagrange dual function. In particular, we have the simple but important inequality $$d^{\star} \leqslant p^{\star}$$
This property is called weak duality. If the equality $d^{\star} = p^{\star}$ holds, i.e., the optimal duality gap is zero, then we say that strong duality holds. If strong duality holds and a dual optimal solution $(\lambda^{\star}, \nu^{\star})$ exists, then any primal optimal point is also a minimizer
of $L(x, \lambda^{\star}, \nu^{\star})$. This fact sometimes allows us to compute a primal optimal solution from a dual optimal solution.
\par
For the best lower bound that can be obtained from the Lagrange dual function one can solve the following optimization problem
\\
maximize   $g(\lambda, \nu)$ \\
subject to $\lambda\succeq 0$\\
This problem is called the Lagrange dual problem associated with the main problem.
Conditions for the optimality of a convex problem is called Karush-Kuhn-Tucker (KKT) conditions. If $f_{i}$ are convex and $h_{i}$ are
affine, and $\tilde{x}, \tilde{\lambda}, \tilde{\nu}$ are any points that satisfy the KKT conditions
$    f_{i}(\tilde{x})\leq0, \quad  i=1,...,m$\\
$    h_{i}(\tilde{x})   =0, \quad  i=1,...,p$\\
$    \tilde{\lambda}_{i} \geq 0, \quad  i=1,...,m$\\
$    \tilde{\lambda}_{i} f_{i}(\tilde{x}) = 0, \quad  i=1,...,m$\\
$    \nabla f_{0}(\tilde{x}) + \sum_{i=1}^{m}\tilde{\lambda}_{i} \nabla f_{i}(\tilde{x})+\sum_{i=1}^{p}\tilde{\nu}_{i} \nabla h_{i}(\tilde{x})=0$\\
then $\tilde{x}$ and $(\tilde{\lambda}, \tilde{\nu})$ are primal and dual optimal, with zero duality gap. In other words, for any convex optimization problem with differentiable objective and constraint functions, any points that satisfy the KKT conditions are primal and dual optimal, and have zero duality gap. Hence, $f_{0}(\tilde{x})= g(\tilde{\lambda}; \tilde{\nu})$.
\\
The condition $\tilde{\lambda}_{i} f_{i}(\tilde{x}) = 0, \quad  i=1,...,m$ is known as complementary slackness; it holds for any primal optimal $\tilde{x}$ and any dual optimal $\tilde{\lambda}, \tilde{\nu}$ (when strong duality holds)

\newpage
\vspace{1cm}\setcounter{section}{0}
\setcounter{equation}{0}
\renewcommand{\theequation}{A-\roman{equation}}
  {\Large{Appendix B: Proving FR inequalities}}\\
\emph{\textbf{a). FR inequality (\ref{LiConst}) of example 1 }}
\par
Here we prove inequality (\ref{LiConst}) for $i_{1}=...=i_{5}=0$. The other cases could be proved similarly.
We use the abbreviations
\begin{equation}\label{}
    \begin{array}{c}
       Tr(\sigma_{i}^{(1)} \ | \alpha\rangle\langle\alpha|)=a_{_{i}} \\
       Tr(\sigma_{i}^{(2)} \ | \beta\rangle\langle\beta|)=b_{_{i}}  \\
       Tr(\sigma_{i}^{(3)} \ | \gamma\rangle\langle\gamma|)=c_{_{i}}.\\
     \end{array}
\end{equation}
where the superscripts $1,2,3$ in  $\sigma_{i}$, denotes the first, second, and third party respectively.
Since $a_{_{1}}^{2}+a_{_{2}}^{2}+a_{_{3}}^{2}=1$ and also the
similar relations hold for $b_{_{i}}$'s and $c_{_{i}}$'s, so the
points $a,b,c$ lie on a unit sphere and we can parameterize their
coordinates by using spherical coordinates $\theta$ and $\varphi$
as follows
$$
\begin{array}{c}
  a_{_{1}}=\sin{\theta_{_{1}}}\cos{\varphi_{_{1}}},  \quad
  a_{_{2}}=\sin{\theta_{_{1}}}\sin{\varphi_{_{1}}},\quad a_{_{3}}=\cos{\theta_{_{1}}} \\
   b_{_{1}}=\sin{\theta_{_{2}}}\cos{\varphi_{_{2}}},  \quad
  b_{_{2}}=\sin{\theta_{_{2}}}\sin{\varphi_{_{2}}},\quad b_{_{3}}=\cos{\theta_{_{2}}} \\
   c_{_{1}}=\sin{\theta_{_{3}}}\cos{\varphi_{_{3}}},  \quad
  c_{_{2}}=\sin{\theta_{_{3}}}\sin{\varphi_{_{3}}},\quad c_{_{3}}=\cos{\theta_{_{3}}}. \\
\end{array}
$$
Now
$$
P_{1}+P_{2}+P_{3}+P_{4}+P_{5}-P_{6}=
$$
$$
[ a_{1}( b_{1} c_{1}+ b_{2} c_{2})+a_{2}( b_{1} c_{3}+ b_{3} c_{2})+a_{3}( b_{2} c_{3}- b_{3} c_{1}) ] \leqslant [1-(b_{1} c_{2}-b_{2} c_{1}-b_{3} c_{3})^{2}]
$$
which is equal or less than one. In the last step we use the Cauchy-Schwartz inequality.
\par
\emph{\textbf{b). FR inequality (\ref{e1}) of example 2 }}
\\
From definition $P_{i}=Tr(Q_{i} \rho_{s})$ we have
$$
\begin{array}{c}
  P_{1}=\cos(\theta_{1}) \sin (\theta_{2}) \sin (\theta_{3}) \cos(\varphi_{2}-\varphi_{3}) , \\
  P_{2}=\cos(\varphi_{1}) \sin (\theta_{1}) \sin (\theta_{2}) \sin (\theta_{3}) \cos(\varphi_{2}-\varphi_{3}) , \\
\end{array}
$$
$$
P_{3}=\sin(\varphi_{1}) \sin (\theta_{1}) \sin (\theta_{2}) \sin (\theta_{3}) \cos(\varphi_{2}-\varphi_{3}) ,
$$
$$
P_{4}=\cos(\theta_{1}) \sin (\theta_{2}) \sin (\theta_{3}) \sin(\varphi_{3}-\varphi_{2}) ,
$$
$$
P_{5}=\cos(\varphi_{1}) \sin (\theta_{1}) \sin (\theta_{2}) \sin (\theta_{3}) \sin(\varphi_{3}-\varphi_{2}) ,
$$
$$
P_{6}=\sin(\varphi_{1}) \sin (\theta_{1}) \sin (\theta_{2}) \sin (\theta_{3}) \sin(\varphi_{3}-\varphi_{2}) ,
$$
$$
P_{7}=\sin (\theta_{2}) \cos (\varphi_{2}) \cos(\theta_{3}) ,
$$
$$
P_{8}=\sin (\theta_{2}) \sin (\varphi_{2}) \cos(\theta_{3}) ,
$$
$$
P_{9}\cos(\theta_{2}) .
$$
Now $\sum_{1}^{6} P_{i}^{2}= \sin^{2}(\theta_{2}) \sin^{2}(\theta_{3})$, and $P_{7}^{2}+P_{8}^{2}=\sin^{2}(\theta_{2}) \cos^{2}(\theta_{3})$ so $\sum_{1}^{8} P_{i}^{2}=\sin^{2}(\theta_{2}) $ and finally $\sum_{1}^{9} P_{i}^{2}=1$. This equation defines the surface of hyper-sphere. As we want to determine the region on and inside of this surface then we can write $\sum_{1}^{9} P_{i}^{2}\leqslant1$, which defines the hyper ball.


\end{document}